\documentclass[aps,pre,reprint,groupedaddress]{revtex4-1}

\usepackage{amsmath}
\usepackage{graphicx,psfrag}
\usepackage[caption=false]{subfig}
\usepackage{verbatim}

\begin{document}


\title{Lattice Gas with Molecular Dynamics Collision Operator}


\author{M. Reza Parsa}
\author{Alexander J. Wagner}
\email[]{alexander.wagner@ndsu.edu}
\homepage[]{www.ndsu.edu/pubweb/$\sim$carswagn}
\affiliation{Department of Physics, North Dakota State University, Fargo, North Dakota 58108, USA\\
Program in Materials and Nanotechnology, North Dakota State University, Fargo, North Dakota 58108, USA}


\date{\today}

\begin{abstract}
  We introduce a lattice gas implementation that is based on coarse-graining a Molecular Dynamics (MD) simulation. Such a lattice gas is similar to standard lattice gases, but its collision operator is informed by an underlying MD simulation. This can be considered an optimal lattice gas implementation because it allows for the representation of any system that can be simulated with MD.  We show here that  equilibrium behavior of the popular lattice Boltzmann algorithm is consistent with this optimal lattice gas. This comparison allows us to make a more accurate identification of the expressions for temperature and pressure in lattice Boltzmann simulations which turn out to be related not only to the physical temperature and pressure but also to the lattice discretization. We show that for any spatial discretization we need to choose a particular temporal discretization to recover the lattice Boltzmann equilibrium.
\end{abstract}


\maketitle

\section{Introduction}
Lattice Boltzmann methods are an important computational tool that is most commonly employed to simulate hydrodynamic systems\cite{succi2001lattice}, and it has been adapted to address many complex phenomena from turbulence \cite{teixeira1998incorporating,yu2005dns} over multi-phase and multi-componet flow\cite{swift1995lattice,shan1995multicomponent,he1997theory,briant2004lattice,yuan2006equations} to pore-scale simulations of porous media\cite{tolke2002lattice,kang2006lattice} and simulations of immersed boundaries \cite{dupin2003multi,jansen2011bijels}. It derives its power from an underlying mesoscopic description that ensures exact mass and momentum conservation. The exact physical meaning of the lattice Boltzmann densities, however, remains poorly understood.

The lattice Boltzmann method was derived as a theoretical tool for the analysis of lattice gas methods \cite{mcnamara1988use}. Lattice gas methods consist of particles moving on a lattice with velocities that connect neighboring sites. After the particles have moved a stochastic collision step rearranges the particles. If these collisions conserve both the number of particles and the total momentum there will be a hydrodynamics limit for mass and momentum conservation equation. The introduction of a hexagonal instead of a square lattice by Frisch, Hasslacher and Pomeau \cite{frisch1986lattice} recovered the necessary isotropy to allow the momentum equation to be related to the Navier-Stokes equation.

These lattice gas models had some deficiencies, one unfavorable feature was a large and essentially uncontrolled amount of noise that required a significant amount of averaging. To derive the Navier-Stokes equations from the lattice gas dynamics a theoretical ensemble average was performed, leading to a lattice Boltzmann representation. Higuera then proposed to simulate the ensemble averaged lattice Boltzmann evolution equation directly, and thereby avoid the need to average results of the lattice gas equation \cite{higuera1989boltzmann, higuera1989lattice}. The collision operation of this first lattice Boltzmann method could be mapped one-to-one to the lattice gas and shared some of the positive features of the lattice gas, like the existence of an H-theorem with unconditional stability, and also some of its deficiencies like velocity dependent viscosities.

It was then realized that there existed much more freedom in the choice of the collision operator, and in particular the relaxation towards a local equilibrium function, often called the Bhatnagar-Gross-Krook (BGK) approach, allowed the full recovery of the Navier-Stokes equation to second order \cite{qian1992lattice}.

At this time a second approach to derive the lattice Boltzmann equation directly from the continuous Boltzmann equation with a BGK collision operator gained popularity \cite{he1997theory}. Over the years several different local equilibrium distributions have been proposed, and currently the most popular method is a standard form of a second order expansion in terms of velocities.



Typically these lattice Boltzmann methods are validated by their ability to recover the Navier-Stokes equation. Here, however, we want to establish a relation to an underlying Molecular Dynamics simulation.
For any Molecular Dynamics simulation, we can bin the particles into lattice cells corresponding to the lattice Boltzmann lattice. We can then observe where the particles in cell $x$ migrate to after a time $\Delta t$, and associate these particles with a lattice velocity vector $v_i = x(t)-x(t+\Delta t)$. These particles will collect at their new lattice cells. After another timestep $\Delta t$ these particles are re-distributed to new lattice sites, and can be associated with new lattice velocities.  We call this representation of the MD simulation Molecular-Dynamics-lattice-gas (MDLG). This redistribution can be understood to be an effective MDLG-collision operator. In some very fundamental sense this is the collision operator that the lattice Boltzmann approach is trying to mimic. The purpose of this paper is to understand the physical meanting of the lattice Boltzmann densities in terms of this fundamental MDLG representation. 

The paper is organized as follows: we first introduce a general idea of a lattice gas and then derive a new lattice gas which consists of a coarse-graining of an underlying MD simulation. We then apply this general idea to a specific MD simulation of a Lennard Jones gas in two dimensions. We analyse the equilibrium properties of the associated MDLG method and show that we are able to predict its mathematical form analytically. We then introduce the lattice Boltzmann method and compare the equilibrium properties of the MDLG method to the lattice Boltzmann equilibrium. We show that there are particular choices for the coarsgraining time and space discretization that lead to equilibria that are compatible with the lattice Boltzmann results. 

\section{\label{LG}Lattice Gas}
At its very basis a lattice gas consists of particles, all located on lattice points, that move with  a lattice velocities $v_i$. What we mean by lattice velocity is that if $x$ is a lattice point, so it $x+v_i$. There are $n_i(x,t)$ particles at time $t$ at position $x$ moving with velocity $v_i$. The evolution consists of two steps. A collision step that redistributes the particles at the lattice point $x$ to different velocities:
\begin{equation}
  n_i^*(x,t) = n_i(x,t)+\Xi_i(\{n_j(x,t)\})
  \label{LGcollision}
\end{equation}
where the collision operator $\Xi_i$ is a function of all the particles and their velocities that are located at lattice point $x$ at time $t$. This collision operator will ensure that none of the locally conserved quantities are changed in the collision process. These locally conserved quantities will vary, depending on the desired physical system that one wants to model. In the majority of cases one will ensure mass and momentum conservation. Early lattice gases restricted the number of particles to at most one per velocity $v_i$ at a lattice site, and the velocity vectors all had the same length, ensuring that mass and energy conservation were synonymous. Each conserved quantity will lead to a corresponding hydrodynamic equation. Most applications focused on the fluid flow, and the key hydrodynamic equations to recover were the continuity and Navier-Stokes equations. Energy conservation is often abandoned in favor of an isothermal condition for many practical applications. Local mass and momentum densities are defined as
\begin{align}
  \rho &= \sum_i n_i\\
  \rho u_\alpha &= \sum_i v_{i\alpha} n_i \label{udef}
\end{align}
and the conservation of these quantities then implies
\begin{align}
  \sum_i \Xi_i &=0,\label{masscons}\\
  \sum_i v_{i\alpha} \Xi_i &=0.\label{momentumcons}
\end{align}
For the new kind of lattice gas collision operator proposed in section \ref{MDLG} we will see that mass conservation of Eq. (\ref{masscons}) is indeed fulfilled, but the momentum conservation of Eq. (\ref{momentumcons}) is not strictly obeyed. Because the new algorithm is based on MD, however, the algorithm conserves momentum rigorously. Its representation of momentum through Eq. (\ref{udef}), however, is inexact.

This collision is then followed by a streaming step
\begin{equation}
  n_i(x+v_i,t+1) = n_i^*(x,t)
  \label{LGstreaming}
\end{equation}
where the particles move to the lattice site indicated by the velocity index $i$, \textit{i.e.} they move from $x$ to $x+v_i$. 
The full evolution equation for these densities can then be written as
\begin{equation}
  n_i(x+v_i, t+1) = n_i(x,t)+ \Xi_i.
  \label{fullLG}
\end{equation}
Of course to make this description complete we need to define the collision operator. Originally lattice gases were defined such that there could be at most one particle for each $n_i$ \cite{frisch1986lattice}.  For the purpose of this paper, however, we will make no such restriction. Instead we investigate a collision operator that is defined by an underlying molecular dynamic simulation.


\section{\label{MDLG}Lattice Gas with Molecular Dynamics collision operator}
In principle most systems of interest for a lattice gas (LG) simulation could be simulated using a Molecular Dynamics (MD) approach as well. MD is a standard tool which follows classical particle trajectories for particles interacting with a pair-potential by numerically integrating Newton's equation of motion.

\begin{figure}
  \includegraphics[width=\columnwidth,clip=true]{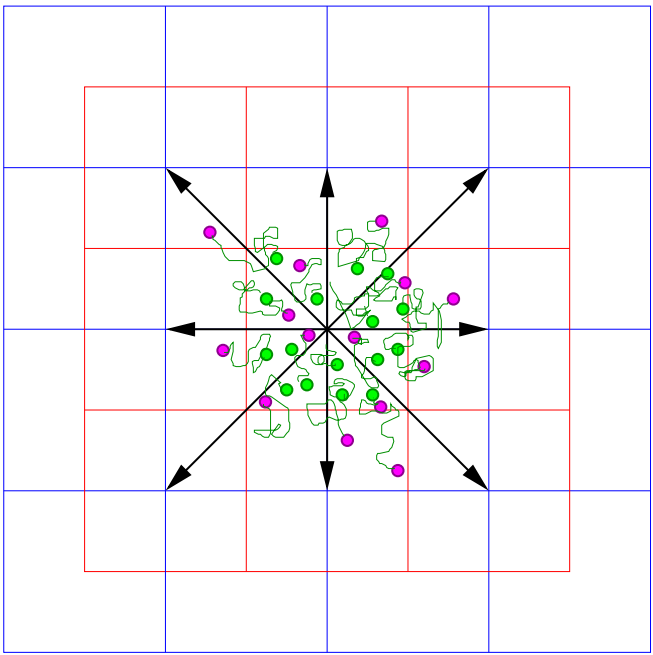}
  \caption{Sketch of the MDLG algorithm: a lattice (blue line) is superimposed on the domain of the MD simulation. Particles in the reciprocal lattice cells (indicated by the red boundaries), are associated with the corresponding lattice point. Particles then get associated with the $n_i$ for the $v_i$ which corresponds to their lattice displacement in the time-interval $\Delta t$.}
  \label{D2Q9MDLGSketch}
\end{figure}

To construct a lattice gas method from a molecular dynamics simulation we overlay a lattice onto our MD simulation. The number of particles in each reciprocal lattice cell around the lattice position $x$ then corresponds to the lattice gas density $\rho(x)$, as shown in Fig. \ref{D2Q9MDLGSketch}. If we then choose a time-step $\Delta t$ we can observe where particles ending up in cell $x$ came from. The number of particles moving from cell $x-v_i$ to cell $x$ then corresponds to the lattice gas occupation number $n_i(x,t)$. What is important to note is that this resulting lattice gas model is fundamentally correct in the sense that will obey the continuity and Navier-Stokes equations simply because the molecular dynamics simulation does so. 

The initial condition for a set of $N$ particles in a finite container with periodic boundary conditions is given by their initial positions $x_i(0)$ and velocities $v_i(0)$. These particles then interact through an interaction pair-potential that we take here to only depend on the distance between the two particles: $V_{ij}=V(|x_i(t)-x_j(t)|)$. The MD simulation then provides (too good accuracy) the trajectories $x_i(t)$ which solve Newton's second law:
\begin{align}
  \frac{dx_i(t)}{dt} &= v_i(t)\\
  \frac{dv_i(t)}{dt} &= -\frac{\partial}{\partial x_i}\left(\frac{1}{2}\sum_{j\neq i} V_{ij}\right)
\end{align}
We now superimpose a lattice onto the computational domain of the MD simulation. For simplicity we can imagine a cubic lattice of lattice spacing $\Delta x$, although any lattice will do here. Let us define a function that determines if a particle resides in a specific cell of the reciprocal lattice associated with a lattice point $x$:
\begin{equation}
  \Delta_x(x')=\left\{\begin{array}{lcl}
  1 & & x_\alpha<x'_{i\alpha}(t)\leq x_\alpha+\Delta x\; \forall \alpha \in \{x,y,z\}\\
  0 & & \mbox{otherwise}
  \end{array}\right.
\end{equation}
Next we pick a time step $\Delta t$. We can now determine the lattice displacement each particle originally residing in a lattice point $x$ experiences. The set of all such displacements makes up the minimal set of lattice velocities $v_i$ for our lattice gas method, and the number of particles associated with this displacement makes up the lattice gas densities $n_i(x,t)$. We define
\begin{equation}
  n_i(x,t) = \sum_j \Delta_x(x_j(t)) \Delta_{x-v_i}(x_j(t-\Delta t)).
  \label{MDni}
\end{equation}
This definition ensures that the particle numbers $n_i(x,t)$ will undergo a streaming step given by Eq. (\ref{LGstreaming}). For any given MD simulation we then know all $n_i(x,t)$.
From Eq. (\ref{fullLG}) we see that the collision operator is then given by
\begin{equation}
  \Xi_i = n_i(x+v_i,t+1)-n_i(x,t).
  \label{eqn:Xi}
\end{equation}
This fully defines the MDLG algorithm, a lattice gas with a collision operator that is defined through an underlying MD simulation. In some sense this is an ideal lattice gas model that can handle even the most complex situations, \textit{i.e.} anything that can be addressed by MD, correctly. The key question is whether this collsion operator can be reduced to some stochastic collision operator that only depends on the local $n_i(x,t)$. Clearly this will only be the case for very simple systems since the MDLB collision operator contains information about temporal and spatial correllations of the underlying MD algorithm and can, in principle, deal with many complex phenomena like liquid-gas-solid coexistence, large varieties of transport parameters, including phenomena at high Knudsen, high Mach, and/or high Reynolds numbers, which we don't expect to be accessible to a simple lattice gas algorithm of Eq. (\ref{fullLG}) with a local collision operator. Such extensions will be subject of future research, but are outside the scope of the current paper.

The local number of particles in lattice cell $x$ at time $t$ is give by
\begin{equation}
  N(x,t) = \sum_j \Delta_x(x_j(t)).
\end{equation}
This is consistent with the lattice gas definition of the local density because
\begin{align}
  N(x,t) &= \sum_i n_i(x,t)\\
  &= \sum_i \sum_j \Delta_x(x_j(t))\Delta_{x-v_i}(x_j(t-\Delta t))\\
  &= \sum_j \Delta_x(x_j(t)).
\end{align}
The last equality follows because
\begin{equation}
  \sum_i \Delta_{x-v_i}(x_j(t-\Delta t)) = 1,
\end{equation}
\textit{i.e.} every particle will be found somewhere on the lattice. Note that we have not yet restricted the velocity set. We will use as many velocities as needed.
Mass conservation of Eq. (\ref{masscons}) is clearly fulfilled since
\begin{align}
  &\sum_i \Xi_i\\
  =& \sum_i [n_i(x+v_i,t+1)-n_i(x,t)]\\
  =& \sum_i \left\{\sum_j \Delta_{x+v_i}[x_j(t+\Delta t)]\Delta_{x}(x_j(t))\right.\nonumber\\
  &\left.-\sum_k \Delta_x(x_k(t))\Delta_{x-v_i}(x_k(t-\Delta t))]\right\}\\
  =& \rho(x,t) - \rho(x,t)\\
  =& 0
  \label{eqn:Mom0}
\end{align}

\begin{figure}
  \includegraphics[width=0.45\columnwidth]{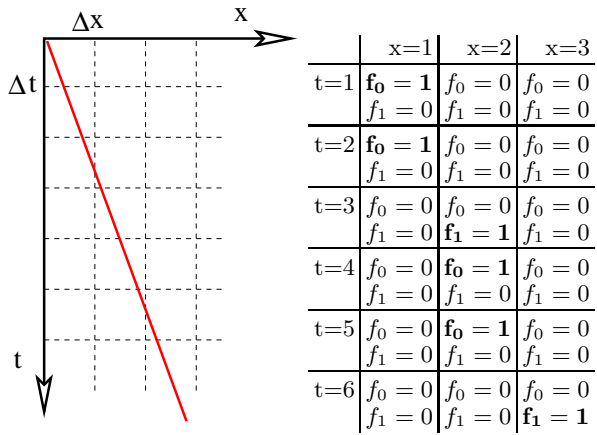}
  \begin{tabular}[b]{r|r|r|r}
    &   x=1 & x=2 & x=3\\
    \hline
    t=1 & $\mathbf{f_0=1}$&$f_0=0$&$f_0=0$\\[-0.1cm]
        & $f_1=0$&$f_1=0$&$f_1=0$\\
    \hline
    t=2 & $\mathbf{f_0=1}$&$f_0=0$&$f_0=0$\\[-0.1cm]
        & $f_1=0$&$f_1=0$&$f_1=0$\\
    \hline
    t=3 & $f_0=0$&$f_0=0$&$f_0=0$\\[-0.1cm]
        & $f_1=0$&$\mathbf{f_1=1}$&$f_1=0$\\
    \hline
    t=4 & $f_0=0$&$\mathbf{f_0=1}$&$f_0=0$\\[-0.1cm]
        & $f_1=0$&$f_1=0$&$f_1=0$\\
    \hline
    t=5 & $f_0=0$&$\mathbf{f_0=1}$&$f_0=0$\\[-0.1cm]
        & $f_1=0$&$f_1=0$&$f_1=0$\\
    \hline
    t=6 & $f_0=0$&$f_0=0$&$f_0=0$\\[-0.1cm]
    & $f_1=0$&$f_1=0$&$\mathbf{f_1=1}$\\
  \end{tabular}
  \caption{Simple thought experiment for lattice gas representation of a particle moving with a constant velocity. Clearly the lattice gas definition of momentum will fluctuate as a function of time even though the underlying MD momentum is conserved. Averaging over all lattice placements, however, will recover the correct momentum.}
\label{fig:1dmom}
\end{figure}

The definition of momentum in the lattice gas sense is typically defined as:
\begin{equation}
  N(x,t) u(x,t) = \sum_i n_i(x,t) v_{i}
\end{equation}
However, relating this to the underlying momentum of the MD simulation is not exact, as can be seen in the example of a single MD particle moving with a velocity less that is not a lattice velocity shown in Fig. \ref{fig:1dmom}. The correspondence could be made exact if we were to introduced an average over all possible placements of the lattice. Such an average would make no difference to the global equilibrium distribution, which is the main focus of the remaining paper. We therefore avoid this additional complication for the current paper. 

Similarly, momentum conservation of Eq. (\ref{momentumcons}) is only exact if we introduce an average over lattice placements:
\begin{align}
  \sum_i v_{i\alpha} \Xi_i \neq 0 \textbf{ in general}
\end{align}
Of course this does not mean that there is a problem with momentum conservation. Instead the problem arises due to the definition of momentum through measured mass transfer between sites for a fixed lattice. 

Despite the apparent lack of momentum conservation the MDLG collision rules are still correct, even without the averaging, since the underlying MD simulation respects momentum conservation.  As such the apparent violation of momentum conservation of the MDLG model is benign. We reserve a closer examination of this averaged lattice gas implementation for a followup paper.

 The key question is then  whether the collision operator (Eq. (\ref{eqn:Xi})) can take the form of Eq. (\ref{LGcollision}), \textit{i.e.} a stochastic collision operator that only depends on the current local occupation numbers $n_j(x,t)$. Since the is a whole ensemble of MD simulations that is consistent with a set of $n_j(x,t)$, and these different MD simulations will lead to different collision terms, it is clear that there can be no exact mapping. However, it is reasonable to hope that we will be able to construct a stochastic lattice gas collision operator that is statistically equivalent to the collision operators for the ensemble of corresponding molecular dynamics simulations. Establishing this is not a trivial task, and we will focus on the easier problem of showing that these collision operators are consistent with the equilibrium behavior of the lattice gas. In the next section we will present the lattice Boltzmann method which conceptually represents the ensemble average of a lattice gas method. Given the complexity of the task we focus in this paper on examining for which, if any, discretizations the MDLG and the standard lattice Boltzmann method give an equivalent equilibrium behavior.

\section{MDLG for an two-dimensional Lennard-Jones gas}
As a test case we use for our underlying MD simulations particles interacting with the standard
Lennard-Jones interaction potential, which is given by 
\begin{equation} V(x)=4\epsilon\left[\left(\frac{\sigma}{x}\right)^{12}-\left(\frac{\sigma}{x}\right)^6\right].
\end{equation}
The interaction strength is controlled by $\epsilon$ and the spatial scaling by $\sigma$.
If $m$ is the mass of a particle we can construct the time-scale
\begin{equation}
  \tau=\sqrt{\frac{m\sigma^2}{\epsilon}}.
\end{equation}
We performed a molecular dynamics simulation using LAMMPS for $100\,000$ particles in a two-dimensional box with the length of $1000\sigma$, corresponding to a nominal volume fraction of 0.0785 if we approximate the particles as circles with diameter $\sigma$. The system was initialized with a homogeneous distribution of particles with a kinetic energy corresponding to a temperature of 50 in the LJ units defined above. This corresponds to a gas at a high temperature, dense enough so that there are a significant number of collisions. The temperature is well above the critical temperature for a liquid-gas coexistence of  $T_c = 1.3120(7)$\cite{potoff1998critical}. We ran an simulation of an equilibrium system with a time-step of $0.0001 \tau$ for $3\,000\,000$ time-steps, \textit{i.e.} up to $\tau=300$. Early time data of $1\,000\,000$ timesteps was discarded, to ensure that we were only probing the equilibrium dynamics.

\begin{figure}
  \includegraphics[width=0.5\columnwidth,clip=true]{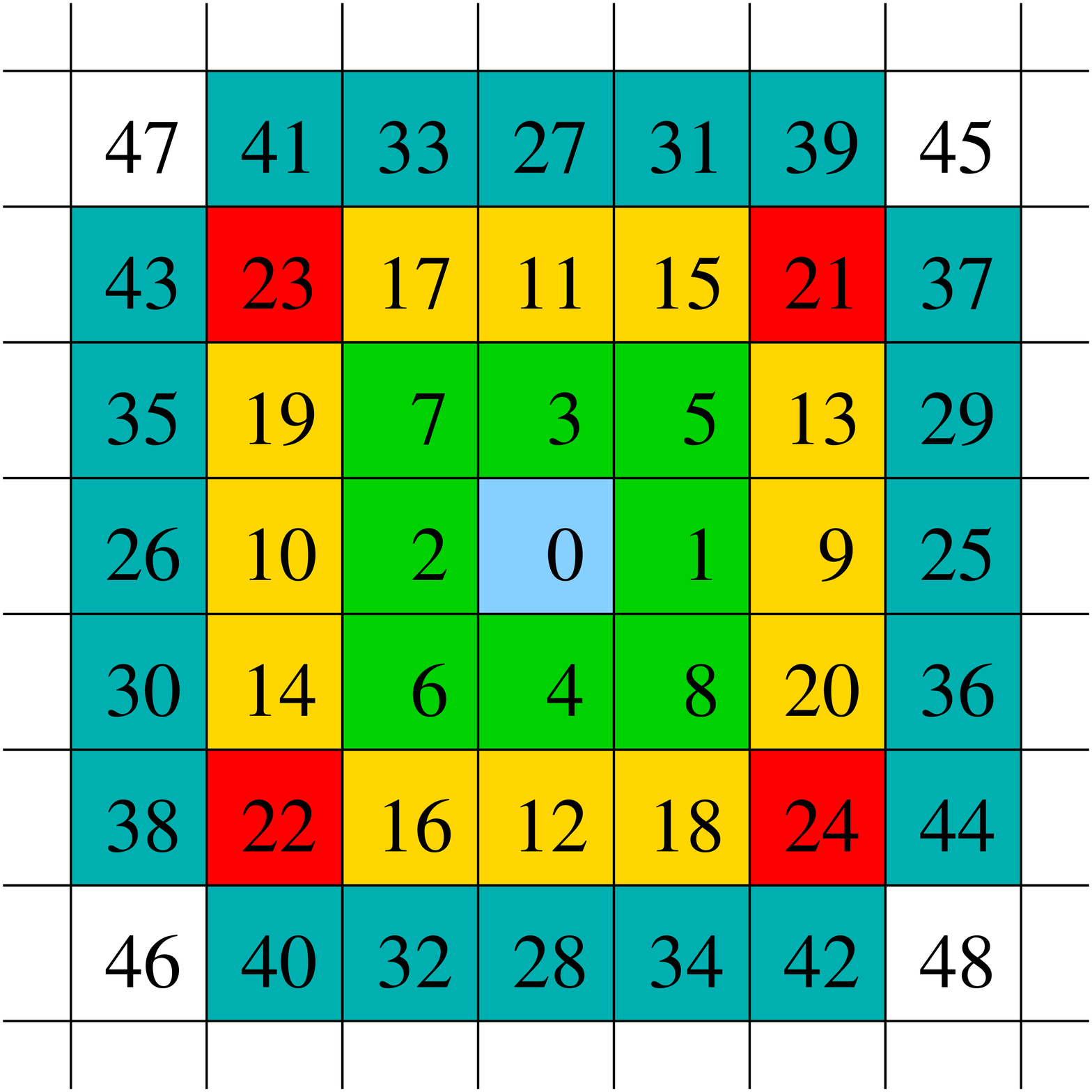}
  \caption{The numbering convention for the velocities $v_i$ in two dimensions. The central point is $0$ and corresponds to velocity $v_0=(0,0)$, and the other velocities are given by the connecting vector between the central point and the lattice point in question.}
  \label{fig:vel}
\end{figure}

We then analyse the resulting MD trajectories to obtain the resulting averaged MDLG occupation numbers $n_i(x,t)$ from Eq. (\ref{MDni}). We should note here that results for different mean velocities $U$ can be obtained simply be using a lattice displaced by $-U\;\Delta t$ for the MDLG analysis. It is therefore not necessary to re-run the MD simulations to examine different mean velocities.

The first information to be gleaned from this is the resulting velocity set for the $v_i$. For small times $\Delta t$, only the nearest neighbors have non-negligible contributions, but as $\Delta t$ is increased more densities get populated. We identify the velocities $v_i$ using the numbering scheme shown in Fig. \ref{fig:vel}. So for $\Delta t\rightarrow 0$ only $v_0-v_8$  will have contributions. These velocities form a complete shell around the central point $v_0$. Most standard lattice Boltzmann methods work hard to make due with this minimal velocity set. This comes at some cost, the most important one is that only one temperature, $\theta=1/3$ in lattice unites, is allowable in lattice units to recover the correct viscous stress tensor (see Eq. (\ref{tet})). For larger $\Delta t$ particle will travel further and a larger velocity set is required. The average occupation numbers are given by the global equilibrium distribution. The next subsection discusses how these equilibrium distribution can be obtained analytically.

\subsection{Global Equilibrium Distribution \label{MDLGequil}}
For a system in thermal equilibrium, sufficient averaging will give an equilibrium distribution
\begin{align}
  f_i^{eq} =& \langle n_i\rangle\\
  =&\sum_j \langle \Delta_x(x_j(t))
  \Delta_{x-v_i}(x_j(t-\Delta t))\rangle.
  \label{fieq}
\end{align}
We can numerically approximate this equilibrium density by averaging the values of the $n_i$ from Eq. (\ref{MDni}) over the whole lattice and for the duration of the simulation. For a given MD simulation the results will depend both on the lattice spacing $\Delta x$ and on the time step $\Delta t$. As mentioned above the MD simulations considered in this paper deal with fairly hot gases, that should be reasibably well approximated by an ideal gas.

To theoretically calculate this expectation value we assume that the particles are uniformly distributed, so only the displacement during the time-interval $\Delta t$ will enter the averaging:
\begin{equation}
  \delta x = x(t)-x(t-\Delta t).
\end{equation}
The key for our averaging will then be the probability of finding such a displacement $P(\delta x)$, and this allows us to write the average as
\begin{align}
  f_i^{eq} =& \frac{\rho^{eq}}{(\Delta x)^d} \int dx \int d(\delta x) \Delta_x(x)\Delta_{x-v_i}(x-\delta x) P(\delta x)\nonumber\\
  =&\frac{\rho^{eq}}{(\Delta x)^d} \int dx \int d(\delta x) \Delta_x(x)\Delta_x(x+v_i-\delta x) P(\delta x)\nonumber\\
  =&\rho^{eq} \int d(\delta x) P(\delta x) W(v_i-\delta x)\nonumber\\
  =&\rho^{eq} \int d(\delta x) P(\delta x+v_i) W(\delta x)
  \label{fieqW}
\end{align}
where $W(x)$ is the d-dimensional wedge function defined as
\begin{align}
  W(x)=&\prod_{\alpha=1}^d W_\alpha(x)\\
  =&\prod_{\alpha=1}^d \left(1-\frac{|x_\alpha|}{\Delta x}\right) \Theta\left(1-\frac{|x_\alpha|}{\Delta x}\right)
\end{align}
where $\Theta$ is the Heaviside function and $\alpha$ denotes cartesian coordinate index.

For very short times $\Delta t$, \textit{i.e.} times shorter than the mean free time between two collisions, particles simply undergo ballistic motion. The velocity distribution of the particles is given by the Maxwell-Boltzmann distribution
\begin{equation}
  P(v) = \frac{1}{(2\pi k_BT)^{d/2}}\exp\left(-\frac{(v-u)^2}{2k_BT}\right).
  \label{MB}
\end{equation}
With this, and neglecting any collisions between the particles, we get for the mean squared displacement in one dimension
\begin{equation}
  \langle (\delta x_\alpha)^2\rangle^{bal} = k_B T (\Delta t)^2.
\end{equation}
The probability distribution for the displacement is then given by
\begin{equation}
  P^{bal}(\delta x) = \frac{1}{(2\pi k_BT)^{d/2}(\Delta t)^d}\exp\left(-\frac{\left(\delta x-u\Delta t\right)^2}{2k_BT(\Delta t)^2}\right).
  \label{MBdx}
\end{equation}
For times much longer than the mean free time particles undergo multiple collisions and instead of following a ballistic motion they will diffuse. If we call the self-diffusion constant $D$ we when have
\begin{equation}
  \langle(\delta x_\alpha)^2\rangle^{dif} =2 D(\Delta t).
\end{equation}
This implies that the probability of the discplacement is given by
\begin{equation}
  P^{diff}(\delta x) = \frac{1}{(4\pi (\Delta t) D)^{d/2}}\exp\left(-\frac{\left(\delta x-u\Delta t\right)^2}{4D(\Delta t)}\right).
\end{equation}
Now since both limiting displacements are given by Gaussian distributions is it reasonable to expect that the intermediate probabilities are also well approximated by a Gaussian and, if we know the mean squared displacement in one dimension $\langle(\delta x_\alpha)^2\rangle$ (and assume isotropy), we get for the probability
\begin{equation}
  P(\delta x) = \frac{1}{(2\pi \langle(\delta x_\alpha)^2\rangle)^{d/2}}\exp\left(-\frac{\left(\delta x-u\Delta t\right)^2}{2\langle(\delta x_\alpha)^2\rangle}\right).
  \label{Pfinal}
\end{equation}
In all of these cases the probability distribution factorizes
\begin{equation}
P(\delta x) = \prod_{\alpha=1}^d P_\alpha(\delta x)
\end{equation}
where
\begin{equation}
  P_\alpha(\delta x) = \frac{1}{\sqrt{2\pi \langle(\delta x_\alpha)^2\rangle}}
  \exp\left(-\frac{(\delta x_\alpha - u_\alpha \Delta t)^2}{2\langle(\delta x_\alpha)^2\rangle}\right)
\end{equation}
and we can write Eq. (\ref{fieqW}) as a product of Gaussian integrals:
\begin{equation}
  f_i^{eq}=\rho^{eq} \prod_{\alpha=1}^d \int d(\delta x_\alpha) P(\delta x_\alpha+v_{i\alpha}) W_\alpha\left(\delta x\right).
  \label{fgen}
\end{equation}
The solution is given by
\begin{equation}
  \frac{f_i^{eq}}{\rho^{eq}}=\prod_{\alpha=1}^d f_{i,\alpha}^{eq}
  \label{feqfull}
\end{equation}
where
\begin{align}
  f_{i,\alpha}^{eq} =& N \left(e^{-\frac{(u_{i,\alpha}-1)^2}{2a^2}}-2e^{-\frac{u_{i,\alpha}^2}{2a^2}}+e^{-\frac{(u_{i,\alpha}+1)^2}{2a^2}}\right)  \nonumber\\
&  +\frac{u_{i,\alpha}-1}{2}\left[\operatorname{erf}(\frac{u_{i,\alpha}-1}{\sqrt{2}a})-\operatorname{erf}(\frac{u_{i,\alpha}}{\sqrt{2}a})\right]\nonumber\\
  &  +\frac{u_{i,\alpha}+1}{2}\left[\operatorname{erf}(\frac{u_{i,\alpha}+1}{\sqrt{2}a})-\operatorname{erf}(\frac{u_{i,\alpha}}{\sqrt{2}a})\right]
  \label{fialpha}
\end{align}
where
\begin{align}
  a^2 &= \frac{\langle(\delta x)^2\rangle}{(\Delta x)^2}\label{eqn:a}\\
  N &=\frac{a}{\sqrt{2\pi}}\\
  u_{i,\alpha} &= v_{i\alpha} - u_\alpha
\end{align}
This is a lattice equilibrium distribution function derived from first principles. At first glance it looks different than other lattice equilibrium distributions, and we will examine its relation to know equilibrium distribution functions below.

\begin{figure}
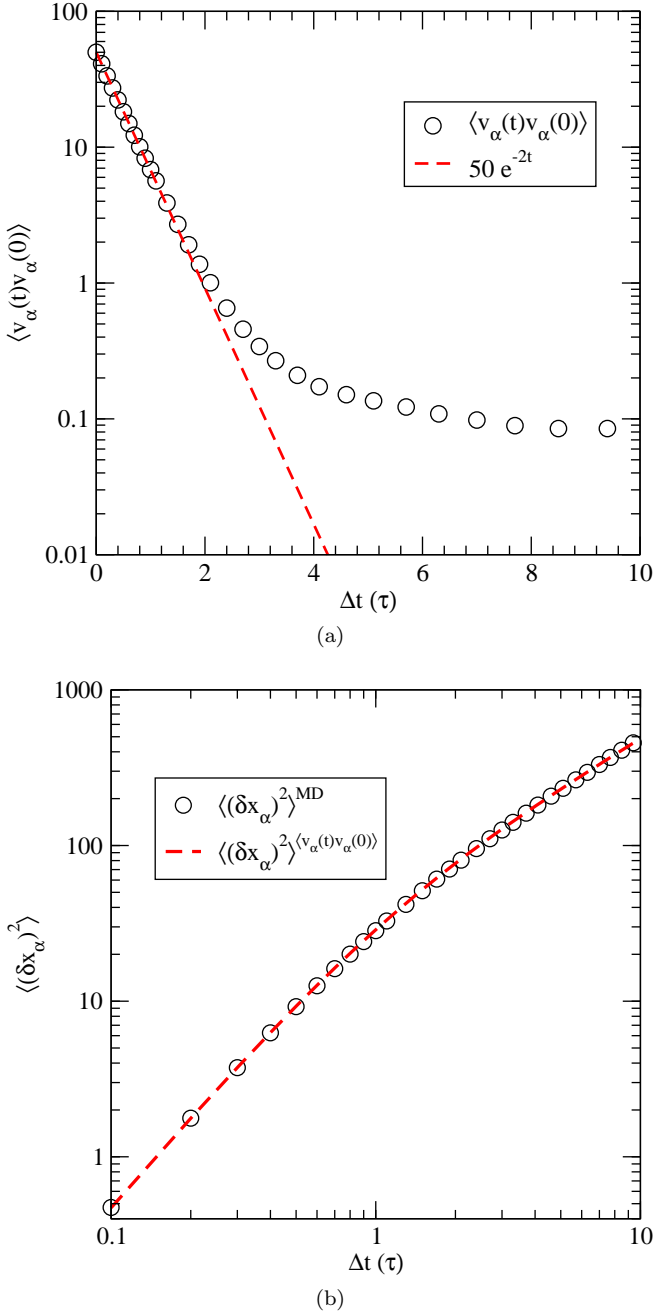

  \subfloat[]{\includegraphics[width=\columnwidth]{V.eps}}\\
  \subfloat[]{\includegraphics[width=\columnwidth]{D.eps}}
  \caption{(a) Measured velocity correlation function from MD simulation data compared to the exponential fit. (b) Measured mean square displacement from MD simulation data compared to the predicted value according to the Eq. (\ref{vc}).}
  \label{fig:VD}
\end{figure}

First we need to fully define the lattice equilibrium distribution. To do so we need to obtain the mean square displacement $\langle(\delta x_\alpha)^2\rangle$. In general the mean square displacement can be measured in our MD simulations, but this would require us to consider a whole function of $\Delta t$ as in input parameter. For our simple ideal gas system we can obtain a simpler dependence on a single parameter by expressing it in terms of the velocity corellation function as
\begin{equation}
  \langle(\delta x_\alpha)^2\rangle=2\int_0^tdt'\;(t-t')\langle v_\alpha(t')v_\alpha(0)\rangle.
  \label{vc}
\end{equation}
For gases this velocity correlation function is typically well approximated by a simple exponential decay. There is also a long range $1/t$ contribution to the velocity correlation function for our two dimensional system, but for the times $\Delta t$ that are of interest here this divergent contribution does not yet contribute noticeably.
\begin{equation}
\langle v_\alpha(t)v_\alpha(0)\rangle= k_BT \exp\left(-\frac{t}{\tau}\right)
\end{equation}
For our system we compare this prediction of an exponentially decaying velocity correlation function to the measured corellation function in Fig. \ref{fig:VD}(a). We see that for early times we see good agreement with this prediction for $\tau = 0.5(0)$. We also see the long time-tale typcical for a two dimensional system, which we ignore here. This is justified below \cite{uhlenbeck1930theory, tsang1977velocity, green1952markoff, green1954markoff, kubo1957statistical, weitz1989nondiffusive}.

Then the mean squared displacement can be predicted according to Eq. (\ref{vc}) as
\begin{align}
\langle(\delta x_\alpha)^2\rangle=
2k_BT \tau^2\bigg(e^{-\frac{t}{\tau}}+\frac{t}{\tau}-1\bigg).\label{vca}
\end{align}
We show that this prediction recovers the measured mean squared displacement well in Fig. \ref{fig:VD}(b). Deviations resulting from the long time tails of the velocity corellation function only show up for later times and larger displacements considered in this paper, which justifies our ignoring these long time tails here.

This fully completes the definition of the MDLG equilibrium function in the case of gases.
To verify our results we compare a numerically measured equilibrium distribution with the theoretically predicted one for different discretizations. The results are shown in Fig. \ref{fig:fi}. The agreement between our theoretical results and the experimental ones is excellent. 

In the next section we introduce the lattice Boltzmann method and then examine the relation of this MDLG equilibrium function with existing lattice equilibrium distribution functions derived for lattice Boltzmann methods.

\section{\label{LB}Lattice Boltzmann}
Lattice Boltzmann methods were derived as ensemble averages of lattice Boltzmann methods. The variables in a lattice Boltzmann method are distribution functions
\begin{equation}
  f_i = \langle n_i\rangle_{neq}
\end{equation}
where the $\langle \cdots \rangle_{neq}$ represents a non-equilibrium ensemble average over microscopic lattice gas states. Taking the same ensemble average, the evolution equation for these lattice Boltzmann densities derives from the underlying lattice gas evolution Eq. (\ref{fullLG})
\begin{equation}
  f_i(x+v_i,t+1) = f_i(x,t)+\Omega_i
\end{equation}
where the collision operator $\Omega_i=\langle \Xi_i\rangle$ is a deterministic function of all the densities at lattice point $x$. We will investigate later if this collision operator can be, at least approximately and for some suitable discretization, be cast in the standard BGK form typically employed for lattice Boltzmann simulations. This question is a crucial first step if one wants to relate lattice Boltzmann to an explicit molecular system which will be represented by our MDLB algorithm.

This standard LB collision operator is a first order BGK approximation and can be written as
\begin{equation}
  \Omega_i = \sum_j \Lambda_{ij} (f_j^0-f_j)
\end{equation}
where $f_j^0$ is a local equilibrium distribution which depends only on the locally conserved quantities
\begin{align}
  \rho &= \sum_i f_i\\
  \rho u_\alpha &= \sum_i v_{i\alpha} f_i.
\end{align}
although other collision operators are also being used \cite{ansumali2002single, ansumali2003minimal, ansumali2005consistent, geier2015cumulant} and it is a longer term goal of the MDLG method to help identify which of these collision operators are most realistic.

To ensure that the lattice Boltzmann equation reproduces the continuity and Navier-Stokes equations in the hydrodynamic limit it is necessary that the equilibrium distribution matches the first four (apart sometimes from a $u^3$ term) velocity moments of the Maxwell-Boltzmann distribution:
\begin{align}
  \sum_i f_i^0 &= \rho \label{mom0}\\
  \sum_i (v_{i\alpha}-u_\alpha) f_i^0 &= 0\\
  \sum_i (v_{i\alpha}-u_\alpha)(v_{i\beta}-u_\beta) f_i^0 &= \rho \theta \delta_{\alpha\beta} \label{tet}\\
  \sum_i (v_{i\alpha}-u_\alpha)(v_{i\beta}-u_\beta)(v_{i\gamma}-u_\gamma)f_i^0 &= Q_{\alpha\beta\gamma}
\label{mom3}
\end{align}
where $Q_{\alpha\beta\gamma}$ should be zero. For velocity sets including only one shell we have $v_{i\alpha}\in\{-1,0,1\}$. For these velocity sets these moments overconstrain the equilibrium distributions. In particular we have $v_{i\alpha}^3=v_{i\alpha}$, which couples the first and the third moment. This is a key source of Galilean invariance violations in lattice Boltzmann\cite{wagner2006investigation}.  These moments can only be reconciled for the special choice of
\begin{equation}
  \theta=1/3
  \label{eqn:13}
\end{equation}
and the third order tensor  $Q_{\alpha\beta\gamma}=\rho u_\alpha u_\beta u_\gamma\ll 1$ which is assumed to be small because $u<0.1$ in typical situations. The equilibrium distribution is typically given in terms of an expansion in terms of the local velocity $u$ up to second order as
\begin{equation}
  f_i^0 = \rho w_i \left(1+\frac{v_{i\alpha}u_\alpha}{\theta}+\frac{1}{2}\frac{v_{i\alpha}u_\alpha v_{i\beta}u_\beta}{\theta^2}-\frac{1}{2} \frac{u_\alpha u_\alpha}{\theta}\right)
\label{eqn:f0lb}
\end{equation}
where the weights $w_i$ depend on the velocity set and summation over repeated Greek indices is implied.
In this article we focus on the question whether this form of an equilibrium distribution is compatible with a concrete MDLB implementation.

This collision operator together with the local equilibrium distribution implies mass and momentum conservation
\begin{align}
  \sum_i \Omega_i &=0,\label{massconsLB}\\
  \sum_i v_{i\alpha} \Omega_i &=0.\label{momentumconsLB}
\end{align}
which is consistent with the typical conditions for lattice gases of Eqs. (\ref{masscons}) and (\ref{momentumcons}).

In the following we will examine the MDLG method for the example of a hot, dilute gas. For this lattice gas we examine the resulting distribution functions and see under which circumstances this lattice gas can reproduce (to some approximation) the lattice Boltzmann method equilibrium distribution Eq. (\ref{eqn:f0lb}).

\section{Relation of MDLG equilibrium functions to lattice Boltzmann equilibria}
We are now in a position to predict for which set of parameters $\Delta x, \Delta t$, if any, we can recover the traditional form of the lattice Boltzmann equilibrium from our MDLB algorithm. Most lattice Boltzmann methods use a limited velocity set that corresponds to a single shell in Fig. \ref{fig:vel}. For our two dimensions this corresponds to 9 velocities. The corresponding equilibrium distribution is typically given as a second order polynomial in the velocities, as we have presented earlier in Eq. (\ref{eqn:f0lb}). For the two dimensional D2Q9 lattice Boltzmann method we consider here the weights $w_i$ in Eq. (\ref{eqn:f0lb}) are given by
\begin{align}
  w_0 &= 4/9\\
  w_{1-4} &= 1/9\\
  w_{5-8} &= 1/36
\end{align}
where the velocity indices correspond to the numbering of Fig. \ref{fig:vel}.

\begin{figure}
  \includegraphics[width=\columnwidth]{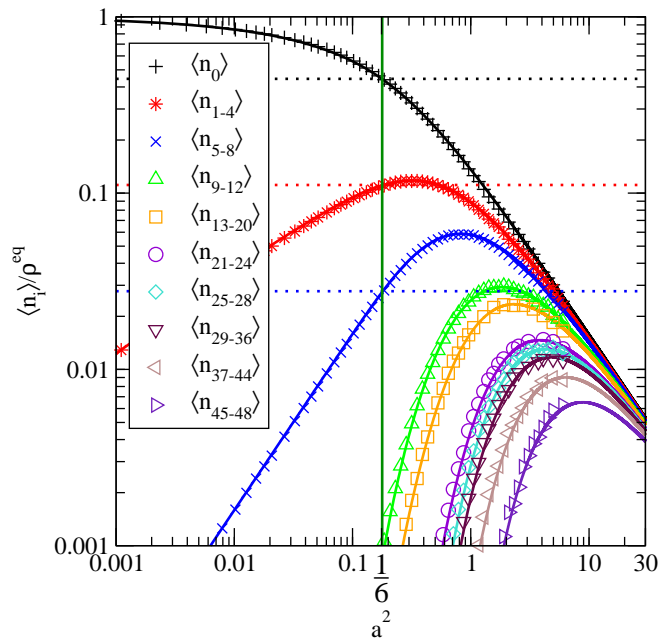}
  \caption{Measured equilibrium distributions $f_i^{eq}$ as a function of the mean-squared displacement measure a from Eq. (\ref{eqn:a}). They are compared to the analytic solution given by Eq. (\ref{feqfull}). We find excellent agreement between the predicted and measured equilibrium distributions. The horizontal lines indicate the value of D2Q9 lattice Boltzmann weights, and the green vertical line indicates the value of $a^2=1/6$ for which these weights agree with the MDLG results.}
  \label{fig:fi}
\end{figure}

In Fig. \ref{fig:fi} we show the lattice Boltzmann weights $w_i$ as horizontal lines. To match the MDLG and LB equilibria we require
\begin{equation}
  f_i^{eq}(a^2)/\rho^{eq}=w_i.
  \label{wif0eqn}
\end{equation}
This is a over-determined system of equations. 

Fortuitously the solutions for the three distinct $w_i$ for a D2Q9 lattice Boltzmann give the same value for $a^2\approx 1/6$ to very good approximation. This is shown as the green vertical line in Fig. \ref{fig:fi}.

\begin{figure}
  \includegraphics[width=\columnwidth]{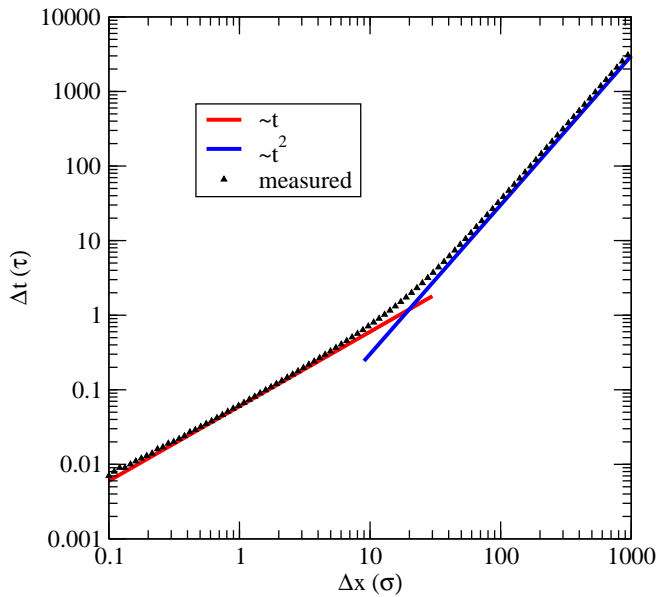}\\
    \caption{Pairs of spatial and temporal discretizations $(\Delta x,\Delta t)$ that lead to equilibrium distributions of MDLG that are consistent with the lattice Boltzmann equilibrium distribution $f_i^0(\rho,0)=w_i$. Notice the two scaling regimes of $\Delta t \propto \Delta x$ for the ballistic regime for small times and the diffusive regime $(\Delta t)^2 \propto \Delta x$ for large times.}
  \label{fig:dtdx}
\end{figure}

This suggests that matching lattice Boltzmann and MD simulations would likely benefit from using the conditions where the $f_i^{eq}$ match up, and the methodology explained above would give guidance on the appropriate time step $\Delta t$ for a given spatial discretization $\Delta x$. Given a $\Delta x$ can numerically solve Eq. (\ref{wif0eqn}) for $\Delta t$ for a system with zero mean momentum.  This is shown in Fig. \ref{fig:dtdx}. We find that there is close agreement between the solutions for different velocities $v_i$. Corresponding to the transition from ballistic to diffusive regime around $\Delta t=\tau$ that we saw in Fig. \ref{fig:VD}(b) we also see a transition here from a $\Delta t \propto t$ regime for $\Delta t\ll 1$ to a $\Delta t \propto t^2$ regime for $\Delta t\gg 1$. We expect this relation that gives $\Delta t$ in terms of $\Delta x$ to be valuable when one tries to generate a coarse-graining transition  between an MD and LG region in a multi-scale numerical method.

\begin{figure}
  \includegraphics[width=\columnwidth]{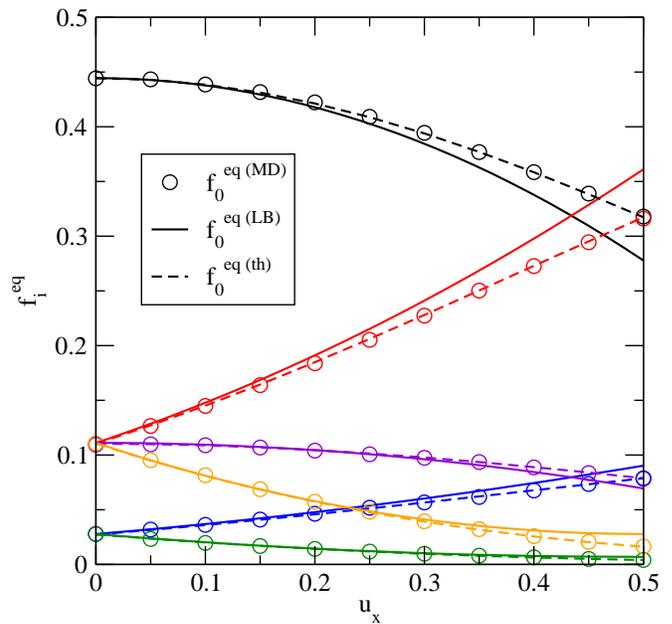}
  \caption{Dependence of the equilibrium distribution function on an imposed velocity $U$ for $\Delta x=100\;\sigma$ and $\Delta t=34.16\;\tau$, corresponding to the green line in Fig. \ref{fig:fi}b. These results are compared to the lattice Boltzmann equilibrium distribution for D2Q9. We find good agreement for velocities below about 0.2.}
  \label{fig:wi_u}
\end{figure}

So far we have only matched the equilibrium distribution at zero velocity. The theory contains the mean velocity as $u$ in Eq. (\ref{fialpha}). For the measurements we could set up different simulations for mean velocities using the algorithm described above in Eq. (\ref{MDni}). It is more practical, however, to move the Grid instead or equivalently use Galilean transformed particle positions of $\hat{x}_i(t)=x_i(t)+u t$ instead. Using this approach to find the equilibrium distribution for different mean velocities $u$ we show our comparison between the measured discrete equilibria (Eq. (\ref{MDni})), their theoretical prediction (Eq. (\ref{feqfull})), and the lattice Boltzmann equilibrium distribution (Eq. (\ref{eqn:f0lb})) in Fig. \ref{fig:wi_u}. For small velocities $|u|<0.1$ we find good agreement between all three densities. This is the relevant range, as lattice Boltzmann is only considered reliable for small enough velocities. The agreement between the measured and predicted MDLG equilibrium distributions continues to be excellent throughout the whole regime. Note, that the agreement would continue to be excellent for larger velocities. We would only need to adapt the velocity set we consider as velocities with magnitude larger than 0.5 are the same as velocities with magnitude larger than 0.5 plus an additional integer lattice displacement. 

\subsection{Moments of the equilibrium distribution}
The key property of the equilibrium distribution in kinetic theory are the velocity moments. For the derivation of the Navier-Stokes equation moments up to third order are required. It is therefore helpful to examine the moments of the discrete MDLG equilibrium distribution, compare them to the expected moment for a lattice Boltzmann equilibrium distribution and examine how these moments relate to the continuous velocity distribution function for the MD simulation.

Let us spend a moment considering these different concepts, since they are not usually clearly separated in a LB derivation. Firstly we have the velocity distribution of the MD simulation, given by Eq. (\ref{MB}). The moments of this velocity distribution are usually used as a rational for constructing LB equilibrium distributions such that the relevant moments of the discrete LB equilibrium distribution function match those of the continuous Maxwell Boltzmann distribution. This is the rational behind the moment Eqs. (\ref{mom0}--\ref{mom3}) for the lattice Boltzmann equilibrium function.

The moments of the discrete MDLG method are not a priori constrained to obey such a constraint, and indeed we don't expect such an agreement for two reasons. First the underlying displacement probabilities (Eq. (\ref{Pfinal})) are in general different from the displacement probabilities (Eq. (\ref{MBdx})) directly related to the Maxwell-Boltzmann distribution (Eq. (\ref{MB})). Second there is no reason to believe that the averaging procedure of Eq. (\ref{fgen}) will preserve the moments in general.

Now let us consider the first three velocity moments specifically. The zeroth moment relates to the total mass. Since the algorithm conserves mass exactly we expect all three approaches to agree on this moment. Indeed, as we saw in Eq. (\ref{eqn:Mom0}) mass is clearly conserved, and consequently this moment will agree for all of the above approach.

The first moment relates to the local momentum. Even if the MDLG approach does not locally conserve the momentum, the averaged momentum of the equilibrium distribution remains exact. This simply follows from the fact that this discrete moment corresponds to the net mass flow through the lattice. Even though this flow can be inexact at any instance in time (since particles may not cross a boundary despite the fact that they are moving) on average the count of particles crossing boundaries has to give the exact mass current.

\begin{figure}
  \includegraphics[width=\columnwidth]{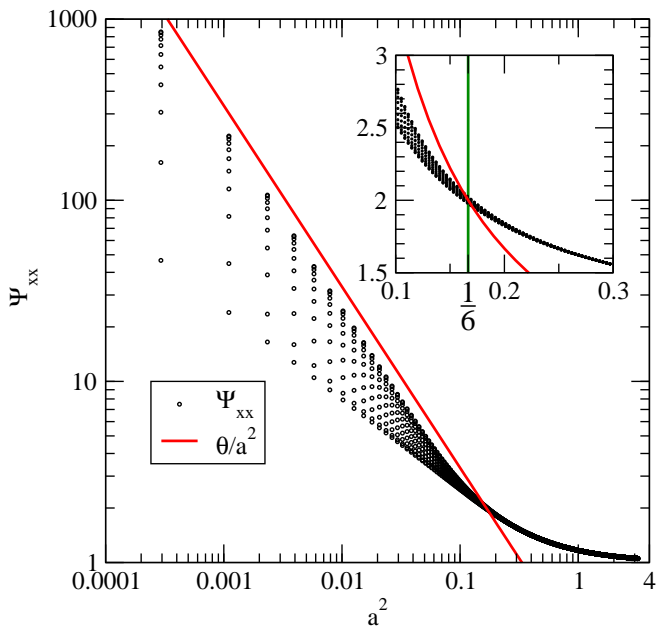}
  \caption{The normalized second moment of Eq. (\ref{eqn:Psi}) as a function of the second moment $a^2$ from Eq. (\ref{eqn:a}) of the mean squared displacement probability. The value of $\Psi$ is shown for different 11 different values of $u_x$ between 0 and 0.5. The values for $u_x=0$ correspond to the bottom points of the graph. For a Galilean-Invariant discrete equilibrium distribution moment the value of $\Psi$ should be independent of $u$.  This is compared to a standard lattice Boltzmann second moment with $\theta=1/3$.}
\label{fig:Psi}
\end{figure}

Let us next consider the second moment. This second moment in the lattice Boltzmann approach (Eq. (\ref{tet})) is related ideal gas equation of state $p=\rho\theta$. We calculate the second moment of our discrete MDLG equilibrium distribution
\begin{equation}
  \Psi_{\alpha\beta}(a,u) = \frac{\sum_i f_i^{eq} (v_{i\alpha}-u_\alpha) (v_{i\beta}-u_\beta)}{\rho^{eq} a^2}.
  \label{eqn:Psi}
\end{equation}
For an equilibrium distribution that obeys the lattice Boltzmann moment (Eq. (\ref{tet})) with a temperature $\theta=a^2$ this expression would give exactly one. For the MDLG equilibrium distribution this second moment is shown in Fig. \ref{fig:Psi}. This shows that the discrete second moment does only agree with the MD temperature for large $a^2\gg 1$. As we see in Fig. \ref{fig:fi} this corresponds to a situation where the populated set of velocities encompasses several shells in Fig. \ref{fig:vel}. For lower values of $a^2$ we find that the second moment diverges. The reason lies in the way we define the discrete equilibrium distribution. Even for very small $\langle (\delta x)^2\rangle $, corresponding to very small $\Delta t$, we will identify a fraction of particles that happen to cross from one lattice point to the next and are therefore assigned a lattice velocity $v_i$ of order one. This appearance of apparent large displacements causes the divergence of the discrete second moment. This effect is significantly enhanced by imposed velocites $u$.   

For $a^2<0.1$ we see that we get significantly diverging values of $\Psi$ for different velocities $u$. This implies that this discrete second moment is not Galilean invariant. It is important to note that despite such violations of Galilean invariance of the discrete moments the full MDLG algorithm does not suffer from a Galilean invariance problem. Instead this is an indication that the collision operator Eq. (\ref{eqn:Xi}) must exactly compensate this apparent Galilean invariance violation. For lattice Boltzmann methods one typically tries to avoid Galilean invariance violations by ensuring that both the local equilibrium distributions and the collision operator independently obey Galilean invariance.

For the lattice Boltzmann method we expect this second moment to be $\Psi_{\alpha\beta}=\theta/a^2 \delta_{\alpha\beta}$. As we saw above [see Eq. (\ref{eqn:13})]lattice Boltzmann methods which have a velocity set consisting of a single shell in velocity space require $\theta = 1/3$. This value of $\theta$ is consistent with the moment of the MDLG equilibrium for a value of $a^2\approx 1/6$. We find $\theta\approx 2 a^2=\langle (\delta x)^2\rangle/(\Delta x)^2$. This corresponds to about the lowest value for $a^2$ where $\Psi$ does not strongly depend on $u$ and would therefore violate Galilean invariance.

\begin{figure}
  \includegraphics[width=\columnwidth]{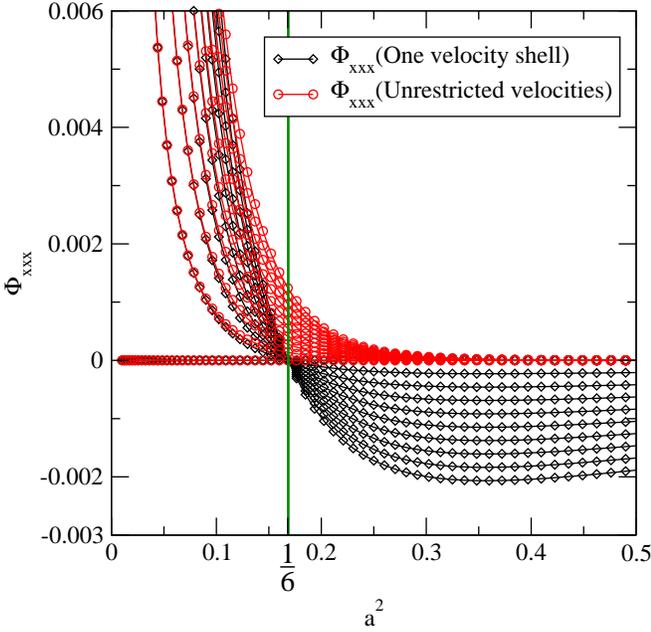}
  \caption{Third velocity moment given by Eq. (\ref{eqn:Phi}) for an unrestricted velocity set and a velocity set restricted to only a single shell of velocities for 10 equally spaced velocities $u_x$ between 0 and 0.001.}
  \label{fig:f3}
\end{figure}

The condition that $\theta=1/3$ came out of a consideration for the third moment for a minimal velocity set with $v_{ix}=v_{ix}^3$. We can define a third moment as
\begin{equation}
  \Phi_{\alpha\beta\gamma} = \frac{\sum_i f_i^{eq} (v_{i\alpha}-u_\alpha) (v_{i\beta}-u_\beta) (v_{i\gamma}-u_\gamma)}{\rho^{eq} a^3}
  \label{eqn:Phi}
\end{equation}
We examine the behavior of $\Phi_{xxx}$ in Fig. \ref{fig:f3}. In fact this third moment should be zero, and for sufficiently large $a$ it converges to zero exponentially. However, if we artificially restrict our velocity set to a single shell, neglecting the small densities for discrete velocities outside the first shell, we find that there is a collapse of $\Phi_{xxx}$ to zero for the same $a^2\approx 1/6$ that we found in Fig. \ref{fig:Psi}. For the full velocity set, however, nothing special occurs at $a^2=1/6$. This initially surprised us because we see in Fig. \ref{fig:fi} that the densities associated with the second shell are still a factor of 30 smaller than the densities of the first shell and might therefore be taken to be negligible. The second shell velocities that are about a factor 2 larger than the first shell velocities, so these densities get multiplied by a factor of $2^3$ which allows these densities to contribute enough that we now have
\begin{equation}
  \sum_i f_i^{eq} v_{ix} \neq \sum_i f_i^{eq} v_{ix}^3
\end{equation}
and they are different enough that the cancellation of terms that is supposed to lead to for $\Phi_{xxx}=0$ at $a^2=1/6$ no longer exists. Instead $\Phi_{xxx}$ monotonously approaches zero.

\begin{figure}
  \includegraphics[width=\columnwidth]{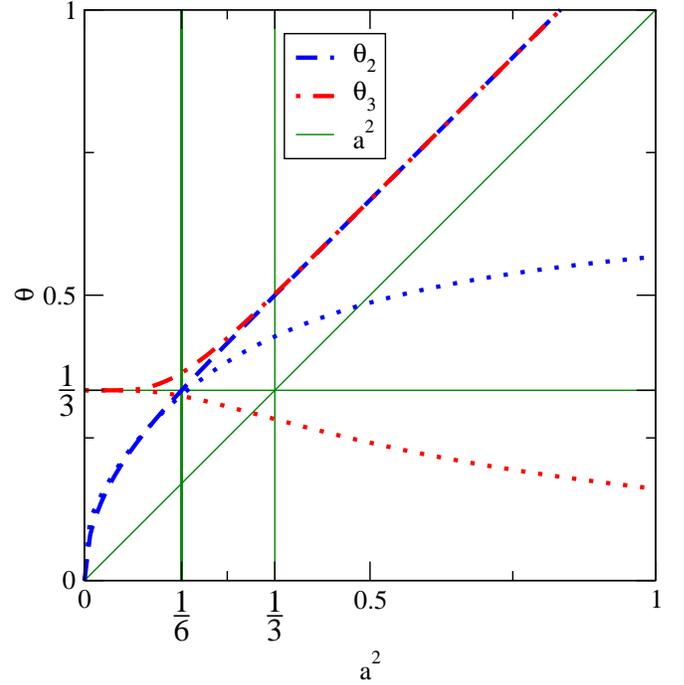}
  \caption{Two manifestations of the temperature $\theta$ from the second and third moments, given by Eqs. (\ref{tet2})  and (\ref{tet3}). We show the moments for the full velocity set and moments for a velocity set restricted to the first shell (thinner lines). For the restricted velocity set we only see agreement between the two moments for $\theta=1/3$ corresponding to $a^2=1/6$.}
  \label{fig:tet}
\end{figure}

The second and third moments of the distribution functions both relate to the lattice Boltzmann temperature of Eq. (\ref{tet}). We can derive two expressions for the temperature:
\begin{align}
  \theta_2 &= a^2 \Psi_{xx}\label{tet2}\\
  \theta_3 &= \frac{1}{3\rho^{eq}}\partial_{u_x} \sum_i f_i^{eq}(\rho^{eq},u) (v_{ix})^3
  \label{tet3}
\end{align}
The dependence of these two quantities on the mean-square displacement is shown in Fig. \ref{fig:tet}. We see that these two definitions only agree with each other for large $a^2$, where agreement is facilitated by utilizing a large set velocity for the discretization of $f_i^{eq}$. For small $a^2$, and therefore a minimal velocity set, we get $\theta_3=1/3$, for the same reason that was discussed before, \textit{i.e.} $v_{ix}=v_{ix}^3$.

If we artificially restrict the velocity set to a single shell (\textit{e.g.} 9 velocities in our two-dimensional example) we affect both definitions of the LB temperature. In this case the agreement for large $a^2$ disappears, and we have exactly one point for $\theta=1/3$ corresponding to $a^2=1/6$ for which both expressions for the temperature agree. This corresponds to the special point in Fig. \ref{fig:f3} where the results for the one-shell velocity set become independent of $u$ (for small enough $u$).

This suggests that there is a serendipitous agreement between the MDLG equilibrium distribution and the standard LB equilibrium distribution for one-shell velocity sets. It allows us to recover velocity moments up to order 3, which is exactly the order required by kinetic theory to recover the continuity and Navier-Stokes equations. As seen in Fig. \ref{fig:Psi} and \ref{fig:f3}, this value is just large enough to avoid apparent Galilean-invariance violations in the moments of the MDLG equilibrium distribution, and just small enough so that the values of the $f_i$ for velocities with $v_{i\alpha}>1$ are small enough not to contribute considerably to the moments of the equilibrium distributions. These effects can only be noticed for the third moment, where they conspire to induce agreement between the measures of temperature implied by the second and third moments, as seen in Fig. \ref{fig:tet}.

\section{Consistent discretizations}
Up to now we have seen how the moments of the equilibrium distribution depend on $a^2$ which is a measure of the mean squared displacement. We saw that the discrete moments of $f_i^{eq}$ differ from the continuous moments of the Maxwell-Boltzmann distribution even in the ballistic regime. The underlying reason for this disagreements result from two conspiring effects. Firstly we only know the position of our particles to lie somewhere within their assigned lattice cell. This uncertainty enters the definition of the $f_i^{eq}$ in Eq. (\ref{fieqW}). Secondly we use a discrete second moment. Here we show how both of these give an offset of $1/12$ in $a^2$ giving rise to a total shift of $1/6$ observed in the previous numerical results.

For simplicity let us consider large times away from the ballistic regime. This occurs without loss of generality if our assumption of a Gaussian distribution in Eq. (\ref{Pfinal}) remains correct. We can then assume the motion of the particles is entirely diffusive with some diffusion constant $D$:
\begin{equation}
\partial_t \rho(x,t) = D \partial_x^2 \rho(x,t)
\end{equation}
If we had known the position of the particle initially, \textit{i.e.} $\rho_c(x,0)=\delta(x)$, then the particle probability density would evolve as
\begin{equation}
\rho_c(x,t) =
  \frac{1}{\sqrt{4\pi Dt}} e^{-\frac{x^2}{4Dt}}
\end{equation}
which has a second moment of
\begin{equation}
  \frac{1}{\sqrt{4\pi Dt}} \int_{-\infty}^\infty x^2 e^{-\frac{x^2}{4 Dt}} dx = 2 Dt
\end{equation}
If we only know that at time $t=0$ the particle is inside a lattice cell centered around the origin with width $\Delta x$, then the density $\rho$ as a function of time is given by
\begin{equation}
  \rho_d(x,t) = \frac{1}{2\Delta x}\left[\mbox{erf}\left(\frac{x+\Delta x/2}{2\sqrt{Dt}}\right)-\mbox{erf}\left(\frac{x-\Delta x/2}{2\sqrt{Dt}}\right)\right]
\end{equation}
which for $t\rightarrow 0$ gives a density that is $1/\Delta x$ inside the interval $[-\Delta x,\Delta x]$ and zero outside it. This probability distribution then spreads out and approaches a Gaussian at late times. 
The second moment of the position is then given by
\begin{equation}
  \int_{-\infty}^\infty x^2\rho_d(x,t)dx = 2Dt+\frac{(\Delta x)^2}{12}
\end{equation}
\textit{i.e.} a simple offsett of $(\Delta x)^2/12$ that does not depend on time.
We can identify
\begin{equation}
  a^2 = \frac{2Dt}{(\Delta x)^2}
\end{equation}
from the definition of Eq. (\ref{eqn:a}).
We then we see that the results are expected to be shifted by $1/12$. However, in Fig. \ref{fig:tet} we clearly see that this only represents half of the observed shift $1/6$.

\begin{figure}
  \includegraphics[width=\columnwidth]{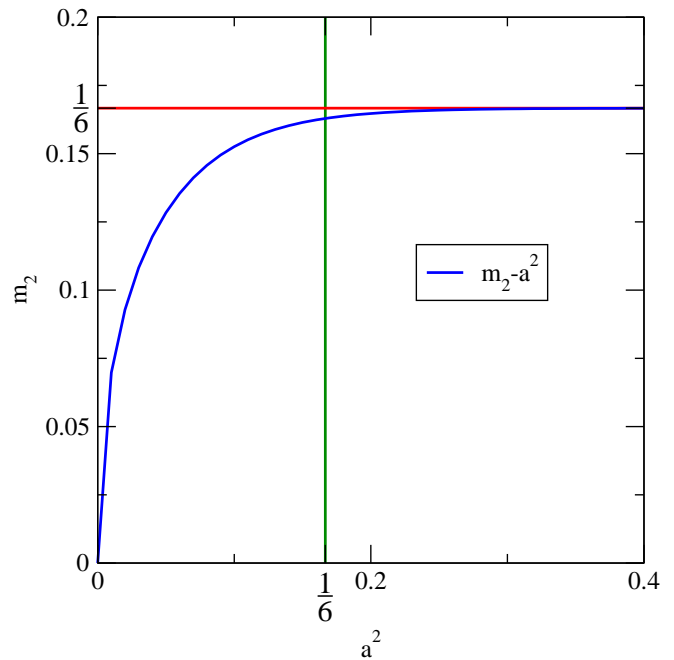}
  \caption{The difference of the discrete second moment $m_2$ of Eq. (\ref{eqn:m2}) minus $a^2$ converges quickly to a constant $1/6$.} 
\label{fig:m2}
\end{figure}

The second effect relates to discretizing the position into displacement bins. We now calculate the discrete second moment (normalized by $(\Delta x)^2$)as
\begin{equation}
  m_2= \sum_{i=-\infty}^\infty  i^2 \int_{i\Delta x-\frac{\Delta x}{2}}^{i\Delta x+\frac{\Delta x}{2}}\rho_d(x,t)\;dx
  \label{eqn:m2}
\end{equation}
In Fig. \ref{fig:m2} we show that this discrete moment quickly converges to $a^2+1/6$. We clearly see that the missing additional offset of $1/12$ that we observed in Fig. \ref{fig:tet} is the result of taking the discrete moment.

This shows that there are two effects of discretization. Firstly the initially broader distribution of particles confined to a lattice site rather than a point shifts the second moment by exactly 1/12. The second effect of discretization is more complicated, particularly for small $a^2$. But for $a^2>0.2$ this effect quickly converges to another offset in $a^2$ of $1/12$. Together they make up the offset of $1/6$, seen repeatedly in our numerical results of Figs. \ref{fig:fi}, \ref{fig:Psi}, \ref{fig:f3}, and \ref{fig:tet}.


\section{Outlook}
In this paper we have introduced a new tool for comparing the results of Molecular Dynamics simulations with those of coarse-grained lattice gas or lattice Boltzmann methods. It consists of re-interpreting the MD results as a lattice gas which we call the MDLG. The dynamics of this special kind of lattice gas is entirely given by the MD simulation, and therefore will be able to give a coarse-grained picture for any results that are obtainable with MD simulations.

Here we focus entirely on the averaged equilibrium behavior and show that there exists a close connection between the equilibria of lattice Boltzmann methods and the equilibrium for the MDLG method, when applied to a hot dilute gas. We were able to determine this equilibrium distribution analytically and were able to verify this analytical solution with the results of the MDLG method. Importantly there is a surprisingly good agreement between our equilibrium distribution and the standard lattice Boltzmann result for carefully chosen (and analytically known) pairs of time and space discretizations $\Delta t$ and $\Delta x$. We were able to understand the observed offset of $1/6$ in the dimensionless measure of the mean squared displacement $a^2$ in terms of our discretization procedure.

This opens the way for a more careful analysis of the fundamental underpinnings of lattice gas and lattice Boltzmann methods. We intend to utilize our MDLG method to investigate the fundamental properties of the collision operator, including its fluctuating properties. Further down the line we hope to investigate how the behavior of liquids  alters the behavior of the MDLG method and examine if MDLG can also be matched with lattice Boltzmann methods. We anticipate that this method will also be instrumental in putting lattice Boltzmann methods for non-ideal and multi-component systems on a firmer footing.

\bibliography{MDLBbib}

\begin{thebibliography}{29}%
\makeatletter
\providecommand \@ifxundefined [1]{%
 \@ifx{#1\undefined}
}%
\providecommand \@ifnum [1]{%
 \ifnum #1\expandafter \@firstoftwo
 \else \expandafter \@secondoftwo
 \fi
}%
\providecommand \@ifx [1]{%
 \ifx #1\expandafter \@firstoftwo
 \else \expandafter \@secondoftwo
 \fi
}%
\providecommand \natexlab [1]{#1}%
\providecommand \enquote  [1]{``#1''}%
\providecommand \bibnamefont  [1]{#1}%
\providecommand \bibfnamefont [1]{#1}%
\providecommand \citenamefont [1]{#1}%
\providecommand \href@noop [0]{\@secondoftwo}%
\providecommand \href [0]{\begingroup \@sanitize@url \@href}%
\providecommand \@href[1]{\@@startlink{#1}\@@href}%
\providecommand \@@href[1]{\endgroup#1\@@endlink}%
\providecommand \@sanitize@url [0]{\catcode `\\12\catcode `\$12\catcode
  `\&12\catcode `\#12\catcode `\^12\catcode `\_12\catcode `\%12\relax}%
\providecommand \@@startlink[1]{}%
\providecommand \@@endlink[0]{}%
\providecommand \url  [0]{\begingroup\@sanitize@url \@url }%
\providecommand \@url [1]{\endgroup\@href {#1}{\urlprefix }}%
\providecommand \urlprefix  [0]{URL }%
\providecommand \Eprint [0]{\href }%
\providecommand \doibase [0]{http://dx.doi.org/}%
\providecommand \selectlanguage [0]{\@gobble}%
\providecommand \bibinfo  [0]{\@secondoftwo}%
\providecommand \bibfield  [0]{\@secondoftwo}%
\providecommand \translation [1]{[#1]}%
\providecommand \BibitemOpen [0]{}%
\providecommand \bibitemStop [0]{}%
\providecommand \bibitemNoStop [0]{.\EOS\space}%
\providecommand \EOS [0]{\spacefactor3000\relax}%
\providecommand \BibitemShut  [1]{\csname bibitem#1\endcsname}%
\let\auto@bib@innerbib\@empty
\bibitem [{\citenamefont {Succi}(2001)}]{succi2001lattice}%
  \BibitemOpen
  \bibfield  {author} {\bibinfo {author} {\bibfnamefont {S.}~\bibnamefont
  {Succi}},\ }\href@noop {} {\emph {\bibinfo {title} {The lattice Boltzmann
  equation: for fluid dynamics and beyond}}}\ (\bibinfo  {publisher} {Oxford
  university press},\ \bibinfo {year} {2001})\BibitemShut {NoStop}%
\bibitem [{\citenamefont {Teixeira}(1998)}]{teixeira1998incorporating}%
  \BibitemOpen
  \bibfield  {author} {\bibinfo {author} {\bibfnamefont {C.~M.}\ \bibnamefont
  {Teixeira}},\ }\href@noop {} {\bibfield  {journal} {\bibinfo  {journal}
  {International Journal of Modern Physics C}\ }\textbf {\bibinfo {volume}
  {9}},\ \bibinfo {pages} {1159} (\bibinfo {year} {1998})}\BibitemShut
  {NoStop}%
\bibitem [{\citenamefont {Yu}\ \emph {et~al.}(2005)\citenamefont {Yu},
  \citenamefont {Girimaji},\ and\ \citenamefont {Luo}}]{yu2005dns}%
  \BibitemOpen
  \bibfield  {author} {\bibinfo {author} {\bibfnamefont {H.}~\bibnamefont
  {Yu}}, \bibinfo {author} {\bibfnamefont {S.~S.}\ \bibnamefont {Girimaji}}, \
  and\ \bibinfo {author} {\bibfnamefont {L.-S.}\ \bibnamefont {Luo}},\
  }\href@noop {} {\bibfield  {journal} {\bibinfo  {journal} {Journal of
  Computational Physics}\ }\textbf {\bibinfo {volume} {209}},\ \bibinfo {pages}
  {599} (\bibinfo {year} {2005})}\BibitemShut {NoStop}%
\bibitem [{\citenamefont {Swift}\ \emph {et~al.}(1995)\citenamefont {Swift},
  \citenamefont {Osborn},\ and\ \citenamefont {Yeomans}}]{swift1995lattice}%
  \BibitemOpen
  \bibfield  {author} {\bibinfo {author} {\bibfnamefont {M.~R.}\ \bibnamefont
  {Swift}}, \bibinfo {author} {\bibfnamefont {W.}~\bibnamefont {Osborn}}, \
  and\ \bibinfo {author} {\bibfnamefont {J.}~\bibnamefont {Yeomans}},\
  }\href@noop {} {\bibfield  {journal} {\bibinfo  {journal} {Physical Review
  Letters}\ }\textbf {\bibinfo {volume} {75}},\ \bibinfo {pages} {830}
  (\bibinfo {year} {1995})}\BibitemShut {NoStop}%
\bibitem [{\citenamefont {Shan}\ and\ \citenamefont
  {Doolen}(1995)}]{shan1995multicomponent}%
  \BibitemOpen
  \bibfield  {author} {\bibinfo {author} {\bibfnamefont {X.}~\bibnamefont
  {Shan}}\ and\ \bibinfo {author} {\bibfnamefont {G.}~\bibnamefont {Doolen}},\
  }\href@noop {} {\bibfield  {journal} {\bibinfo  {journal} {Journal of
  Statistical Physics}\ }\textbf {\bibinfo {volume} {81}},\ \bibinfo {pages}
  {379} (\bibinfo {year} {1995})}\BibitemShut {NoStop}%
\bibitem [{\citenamefont {He}\ and\ \citenamefont {Luo}(1997)}]{he1997theory}%
  \BibitemOpen
  \bibfield  {author} {\bibinfo {author} {\bibfnamefont {X.}~\bibnamefont
  {He}}\ and\ \bibinfo {author} {\bibfnamefont {L.-S.}\ \bibnamefont {Luo}},\
  }\href@noop {} {\bibfield  {journal} {\bibinfo  {journal} {Physical Review
  E}\ }\textbf {\bibinfo {volume} {56}},\ \bibinfo {pages} {6811} (\bibinfo
  {year} {1997})}\BibitemShut {NoStop}%
\bibitem [{\citenamefont {Briant}\ \emph {et~al.}(2004)\citenamefont {Briant},
  \citenamefont {Wagner},\ and\ \citenamefont {Yeomans}}]{briant2004lattice}%
  \BibitemOpen
  \bibfield  {author} {\bibinfo {author} {\bibfnamefont {A.}~\bibnamefont
  {Briant}}, \bibinfo {author} {\bibfnamefont {A.}~\bibnamefont {Wagner}}, \
  and\ \bibinfo {author} {\bibfnamefont {J.}~\bibnamefont {Yeomans}},\
  }\href@noop {} {\bibfield  {journal} {\bibinfo  {journal} {Physical Review
  E}\ }\textbf {\bibinfo {volume} {69}},\ \bibinfo {pages} {031602} (\bibinfo
  {year} {2004})}\BibitemShut {NoStop}%
\bibitem [{\citenamefont {Yuan}\ and\ \citenamefont
  {Schaefer}(2006)}]{yuan2006equations}%
  \BibitemOpen
  \bibfield  {author} {\bibinfo {author} {\bibfnamefont {P.}~\bibnamefont
  {Yuan}}\ and\ \bibinfo {author} {\bibfnamefont {L.}~\bibnamefont
  {Schaefer}},\ }\href@noop {} {\bibfield  {journal} {\bibinfo  {journal}
  {Physics of Fluids}\ }\textbf {\bibinfo {volume} {18}},\ \bibinfo {pages}
  {042101} (\bibinfo {year} {2006})}\BibitemShut {NoStop}%
\bibitem [{\citenamefont {T{\"o}lke}(2002)}]{tolke2002lattice}%
  \BibitemOpen
  \bibfield  {author} {\bibinfo {author} {\bibfnamefont {J.}~\bibnamefont
  {T{\"o}lke}},\ }\href@noop {} {\bibfield  {journal} {\bibinfo  {journal}
  {Philosophical Transactions of the Royal Society of London A: Mathematical,
  Physical and Engineering Sciences}\ }\textbf {\bibinfo {volume} {360}},\
  \bibinfo {pages} {535} (\bibinfo {year} {2002})}\BibitemShut {NoStop}%
\bibitem [{\citenamefont {Kang}\ \emph {et~al.}(2006)\citenamefont {Kang},
  \citenamefont {Lichtner},\ and\ \citenamefont {Zhang}}]{kang2006lattice}%
  \BibitemOpen
  \bibfield  {author} {\bibinfo {author} {\bibfnamefont {Q.}~\bibnamefont
  {Kang}}, \bibinfo {author} {\bibfnamefont {P.~C.}\ \bibnamefont {Lichtner}},
  \ and\ \bibinfo {author} {\bibfnamefont {D.}~\bibnamefont {Zhang}},\
  }\href@noop {} {\bibfield  {journal} {\bibinfo  {journal} {Journal of
  Geophysical Research: Solid Earth}\ }\textbf {\bibinfo {volume} {111}}
  (\bibinfo {year} {2006})}\BibitemShut {NoStop}%
\bibitem [{\citenamefont {Dupin}\ \emph {et~al.}(2003)\citenamefont {Dupin},
  \citenamefont {Halliday},\ and\ \citenamefont {Care}}]{dupin2003multi}%
  \BibitemOpen
  \bibfield  {author} {\bibinfo {author} {\bibfnamefont {M.}~\bibnamefont
  {Dupin}}, \bibinfo {author} {\bibfnamefont {I.}~\bibnamefont {Halliday}}, \
  and\ \bibinfo {author} {\bibfnamefont {C.}~\bibnamefont {Care}},\ }\href@noop
  {} {\bibfield  {journal} {\bibinfo  {journal} {Journal of Physics A:
  Mathematical and General}\ }\textbf {\bibinfo {volume} {36}},\ \bibinfo
  {pages} {8517} (\bibinfo {year} {2003})}\BibitemShut {NoStop}%
\bibitem [{\citenamefont {Jansen}\ and\ \citenamefont
  {Harting}(2011)}]{jansen2011bijels}%
  \BibitemOpen
  \bibfield  {author} {\bibinfo {author} {\bibfnamefont {F.}~\bibnamefont
  {Jansen}}\ and\ \bibinfo {author} {\bibfnamefont {J.}~\bibnamefont
  {Harting}},\ }\href@noop {} {\bibfield  {journal} {\bibinfo  {journal}
  {Physical Review E}\ }\textbf {\bibinfo {volume} {83}},\ \bibinfo {pages}
  {046707} (\bibinfo {year} {2011})}\BibitemShut {NoStop}%
\bibitem [{\citenamefont {McNamara}\ and\ \citenamefont
  {Zanetti}(1988)}]{mcnamara1988use}%
  \BibitemOpen
  \bibfield  {author} {\bibinfo {author} {\bibfnamefont {G.~R.}\ \bibnamefont
  {McNamara}}\ and\ \bibinfo {author} {\bibfnamefont {G.}~\bibnamefont
  {Zanetti}},\ }\href@noop {} {\bibfield  {journal} {\bibinfo  {journal}
  {Physical review letters}\ }\textbf {\bibinfo {volume} {61}},\ \bibinfo
  {pages} {2332} (\bibinfo {year} {1988})}\BibitemShut {NoStop}%
\bibitem [{\citenamefont {Frisch}\ \emph {et~al.}(1986)\citenamefont {Frisch},
  \citenamefont {Hasslacher},\ and\ \citenamefont
  {Pomeau}}]{frisch1986lattice}%
  \BibitemOpen
  \bibfield  {author} {\bibinfo {author} {\bibfnamefont {U.}~\bibnamefont
  {Frisch}}, \bibinfo {author} {\bibfnamefont {B.}~\bibnamefont {Hasslacher}},
  \ and\ \bibinfo {author} {\bibfnamefont {Y.}~\bibnamefont {Pomeau}},\
  }\href@noop {} {\bibfield  {journal} {\bibinfo  {journal} {Physical review
  letters}\ }\textbf {\bibinfo {volume} {56}},\ \bibinfo {pages} {1505}
  (\bibinfo {year} {1986})}\BibitemShut {NoStop}%
\bibitem [{\citenamefont {Higuera}\ and\ \citenamefont
  {Jimenez}(1989)}]{higuera1989boltzmann}%
  \BibitemOpen
  \bibfield  {author} {\bibinfo {author} {\bibfnamefont {F.}~\bibnamefont
  {Higuera}}\ and\ \bibinfo {author} {\bibfnamefont {J.}~\bibnamefont
  {Jimenez}},\ }\href@noop {} {\bibfield  {journal} {\bibinfo  {journal} {EPL
  (Europhysics Letters)}\ }\textbf {\bibinfo {volume} {9}},\ \bibinfo {pages}
  {663} (\bibinfo {year} {1989})}\BibitemShut {NoStop}%
\bibitem [{\citenamefont {Higuera}\ \emph {et~al.}(1989)\citenamefont
  {Higuera}, \citenamefont {Succi},\ and\ \citenamefont
  {Benzi}}]{higuera1989lattice}%
  \BibitemOpen
  \bibfield  {author} {\bibinfo {author} {\bibfnamefont {F.}~\bibnamefont
  {Higuera}}, \bibinfo {author} {\bibfnamefont {S.}~\bibnamefont {Succi}}, \
  and\ \bibinfo {author} {\bibfnamefont {R.}~\bibnamefont {Benzi}},\
  }\href@noop {} {\bibfield  {journal} {\bibinfo  {journal} {EPL (Europhysics
  Letters)}\ }\textbf {\bibinfo {volume} {9}},\ \bibinfo {pages} {345}
  (\bibinfo {year} {1989})}\BibitemShut {NoStop}%
\bibitem [{\citenamefont {Qian}\ \emph {et~al.}(1992)\citenamefont {Qian},
  \citenamefont {d'Humi{\`e}res},\ and\ \citenamefont
  {Lallemand}}]{qian1992lattice}%
  \BibitemOpen
  \bibfield  {author} {\bibinfo {author} {\bibfnamefont {Y.}~\bibnamefont
  {Qian}}, \bibinfo {author} {\bibfnamefont {D.}~\bibnamefont
  {d'Humi{\`e}res}}, \ and\ \bibinfo {author} {\bibfnamefont {P.}~\bibnamefont
  {Lallemand}},\ }\href@noop {} {\bibfield  {journal} {\bibinfo  {journal} {EPL
  (Europhysics Letters)}\ }\textbf {\bibinfo {volume} {17}},\ \bibinfo {pages}
  {479} (\bibinfo {year} {1992})}\BibitemShut {NoStop}%
\bibitem [{\citenamefont {Potoff}\ and\ \citenamefont
  {Panagiotopoulos}(1998)}]{potoff1998critical}%
  \BibitemOpen
  \bibfield  {author} {\bibinfo {author} {\bibfnamefont {J.~J.}\ \bibnamefont
  {Potoff}}\ and\ \bibinfo {author} {\bibfnamefont {A.~Z.}\ \bibnamefont
  {Panagiotopoulos}},\ }\href@noop {} {\bibfield  {journal} {\bibinfo
  {journal} {The Journal of chemical physics}\ }\textbf {\bibinfo {volume}
  {109}},\ \bibinfo {pages} {10914} (\bibinfo {year} {1998})}\BibitemShut
  {NoStop}%
\bibitem [{\citenamefont {Uhlenbeck}\ and\ \citenamefont
  {Ornstein}(1930)}]{uhlenbeck1930theory}%
  \BibitemOpen
  \bibfield  {author} {\bibinfo {author} {\bibfnamefont {G.~E.}\ \bibnamefont
  {Uhlenbeck}}\ and\ \bibinfo {author} {\bibfnamefont {L.~S.}\ \bibnamefont
  {Ornstein}},\ }\href@noop {} {\bibfield  {journal} {\bibinfo  {journal}
  {Physical review}\ }\textbf {\bibinfo {volume} {36}},\ \bibinfo {pages} {823}
  (\bibinfo {year} {1930})}\BibitemShut {NoStop}%
\bibitem [{\citenamefont {Tsang}\ and\ \citenamefont
  {Tang}(1977)}]{tsang1977velocity}%
  \BibitemOpen
  \bibfield  {author} {\bibinfo {author} {\bibfnamefont {T.}~\bibnamefont
  {Tsang}}\ and\ \bibinfo {author} {\bibfnamefont {H.}~\bibnamefont {Tang}},\
  }\href@noop {} {\bibfield  {journal} {\bibinfo  {journal} {Physical Review
  A}\ }\textbf {\bibinfo {volume} {15}},\ \bibinfo {pages} {1696} (\bibinfo
  {year} {1977})}\BibitemShut {NoStop}%
\bibitem [{\citenamefont {Green}(1952)}]{green1952markoff}%
  \BibitemOpen
  \bibfield  {author} {\bibinfo {author} {\bibfnamefont {M.~S.}\ \bibnamefont
  {Green}},\ }\href@noop {} {\bibfield  {journal} {\bibinfo  {journal} {The
  Journal of Chemical Physics}\ }\textbf {\bibinfo {volume} {20}},\ \bibinfo
  {pages} {1281} (\bibinfo {year} {1952})}\BibitemShut {NoStop}%
\bibitem [{\citenamefont {Green}(1954)}]{green1954markoff}%
  \BibitemOpen
  \bibfield  {author} {\bibinfo {author} {\bibfnamefont {M.~S.}\ \bibnamefont
  {Green}},\ }\href@noop {} {\bibfield  {journal} {\bibinfo  {journal} {The
  Journal of Chemical Physics}\ }\textbf {\bibinfo {volume} {22}},\ \bibinfo
  {pages} {398} (\bibinfo {year} {1954})}\BibitemShut {NoStop}%
\bibitem [{\citenamefont {Kubo}(1957)}]{kubo1957statistical}%
  \BibitemOpen
  \bibfield  {author} {\bibinfo {author} {\bibfnamefont {R.}~\bibnamefont
  {Kubo}},\ }\href@noop {} {\bibfield  {journal} {\bibinfo  {journal} {Journal
  of the Physical Society of Japan}\ }\textbf {\bibinfo {volume} {12}},\
  \bibinfo {pages} {570} (\bibinfo {year} {1957})}\BibitemShut {NoStop}%
\bibitem [{\citenamefont {Weitz}\ \emph {et~al.}(1989)\citenamefont {Weitz},
  \citenamefont {Pine}, \citenamefont {Pusey},\ and\ \citenamefont
  {Tough}}]{weitz1989nondiffusive}%
  \BibitemOpen
  \bibfield  {author} {\bibinfo {author} {\bibfnamefont {D.}~\bibnamefont
  {Weitz}}, \bibinfo {author} {\bibfnamefont {D.}~\bibnamefont {Pine}},
  \bibinfo {author} {\bibfnamefont {P.}~\bibnamefont {Pusey}}, \ and\ \bibinfo
  {author} {\bibfnamefont {R.}~\bibnamefont {Tough}},\ }\href@noop {}
  {\bibfield  {journal} {\bibinfo  {journal} {Physical review letters}\
  }\textbf {\bibinfo {volume} {63}},\ \bibinfo {pages} {1747} (\bibinfo {year}
  {1989})}\BibitemShut {NoStop}%
\bibitem [{\citenamefont {Ansumali}\ and\ \citenamefont
  {Karlin}(2002)}]{ansumali2002single}%
  \BibitemOpen
  \bibfield  {author} {\bibinfo {author} {\bibfnamefont {S.}~\bibnamefont
  {Ansumali}}\ and\ \bibinfo {author} {\bibfnamefont {I.~V.}\ \bibnamefont
  {Karlin}},\ }\href@noop {} {\bibfield  {journal} {\bibinfo  {journal}
  {Physical Review E}\ }\textbf {\bibinfo {volume} {65}},\ \bibinfo {pages}
  {056312} (\bibinfo {year} {2002})}\BibitemShut {NoStop}%
\bibitem [{\citenamefont {Ansumali}\ \emph {et~al.}(2003)\citenamefont
  {Ansumali}, \citenamefont {Karlin},\ and\ \citenamefont
  {{\"O}ttinger}}]{ansumali2003minimal}%
  \BibitemOpen
  \bibfield  {author} {\bibinfo {author} {\bibfnamefont {S.}~\bibnamefont
  {Ansumali}}, \bibinfo {author} {\bibfnamefont {I.~V.}\ \bibnamefont
  {Karlin}}, \ and\ \bibinfo {author} {\bibfnamefont {H.~C.}\ \bibnamefont
  {{\"O}ttinger}},\ }\href@noop {} {\bibfield  {journal} {\bibinfo  {journal}
  {EPL (Europhysics Letters)}\ }\textbf {\bibinfo {volume} {63}},\ \bibinfo
  {pages} {798} (\bibinfo {year} {2003})}\BibitemShut {NoStop}%
\bibitem [{\citenamefont {Ansumali}\ and\ \citenamefont
  {Karlin}(2005)}]{ansumali2005consistent}%
  \BibitemOpen
  \bibfield  {author} {\bibinfo {author} {\bibfnamefont {S.}~\bibnamefont
  {Ansumali}}\ and\ \bibinfo {author} {\bibfnamefont {I.~V.}\ \bibnamefont
  {Karlin}},\ }\href@noop {} {\bibfield  {journal} {\bibinfo  {journal}
  {Physical review letters}\ }\textbf {\bibinfo {volume} {95}},\ \bibinfo
  {pages} {260605} (\bibinfo {year} {2005})}\BibitemShut {NoStop}%
\bibitem [{\citenamefont {Geier}\ \emph {et~al.}(2015)\citenamefont {Geier},
  \citenamefont {Sch{\"o}nherr}, \citenamefont {Pasquali},\ and\ \citenamefont
  {Krafczyk}}]{geier2015cumulant}%
  \BibitemOpen
  \bibfield  {author} {\bibinfo {author} {\bibfnamefont {M.}~\bibnamefont
  {Geier}}, \bibinfo {author} {\bibfnamefont {M.}~\bibnamefont
  {Sch{\"o}nherr}}, \bibinfo {author} {\bibfnamefont {A.}~\bibnamefont
  {Pasquali}}, \ and\ \bibinfo {author} {\bibfnamefont {M.}~\bibnamefont
  {Krafczyk}},\ }\href@noop {} {\bibfield  {journal} {\bibinfo  {journal}
  {Computers \& Mathematics with Applications}\ }\textbf {\bibinfo {volume}
  {70}},\ \bibinfo {pages} {507} (\bibinfo {year} {2015})}\BibitemShut
  {NoStop}%
\bibitem [{\citenamefont {Wagner}\ and\ \citenamefont
  {Li}(2006)}]{wagner2006investigation}%
  \BibitemOpen
  \bibfield  {author} {\bibinfo {author} {\bibfnamefont {A.}~\bibnamefont
  {Wagner}}\ and\ \bibinfo {author} {\bibfnamefont {Q.}~\bibnamefont {Li}},\
  }\href@noop {} {\bibfield  {journal} {\bibinfo  {journal} {Physica A:
  Statistical Mechanics and its Applications}\ }\textbf {\bibinfo {volume}
  {362}},\ \bibinfo {pages} {105} (\bibinfo {year} {2006})}\BibitemShut
  {NoStop}%
\end{thebibliography}%

\end{document}